\documentstyle[prl,multicol,amsmath,aps,psfig]{revtex}
\begin{document}

\title{ Origin of the roughness exponent in elastic strings at the
depinning threshold} \author{Alberto Rosso and Werner Krauth
\footnote{rosso@lps.ens.fr; krauth@lps.ens.fr,
http://www.lps.ens.fr/$\tilde{\;}$krauth} 
}
\address{CNRS-Laboratoire de Physique Statistique \\
Ecole Normale Sup{\'{e}}rieure,
24, rue Lhomond, 75231 Paris Cedex 05, France}
\maketitle
\begin{abstract} 
Within a recently developed framework of  dynamical Monte Carlo
algorithms, we compute the roughness exponent $\zeta$ of driven
elastic strings at the depinning threshold in $1+1$ dimensions for
different functional forms of the (short-range) elastic energy.
A purely harmonic elastic energy leads to an unphysical value for
$\zeta$.  We include supplementary   terms in the  elastic energy
of at least quartic order in the local extension.  We then find a
roughness exponent of $\zeta \simeq 0.63$,  which coincides with the
one obtained for different cellular automaton models of directed
percolation depinning.  The  quartic term translates into a nonlinear
piece which changes the roughness exponent in the corresponding
continuum equation of motion.  We discuss the implications of our
analysis for higher-dimensional elastic manifolds in disordered
media.

\end{abstract}

\begin{multicols}{2}
\narrowtext
The competition between elasticity and disorder is a central theme
of current research in statistical physics. The pinning of flux
lines in a type-II superconductor \cite{supra}, the motion of a
charge density wave \cite{CDW}, an interface in magnets and several
realizations of surface  growth \cite{Barabasi} are all governed
by these two antagonistic mechanisms, one trying to smooth the
surface, the other striving to distort it.  These two mechanisms
are already at work in the simplest such system, a one-dimensional
elastic string in a two-dimensional medium, which today is far from
being solved.  In this paper, we report progress in our understanding
of the zero-temperature motion of this system at the depinning
threshold, defined by the critical driving force $f_c$. Above $f_c$,
the elastic string flows with finite velocity, while it is pinned
for forces $f \le f_c$.  We consider the problem on the lattice,
but keep full contact with the continuum description.

Previously \cite{RossoKrauth}, we showed that the dynamical Monte
Carlo method can be reconciled with the continuum equation of motion
approach if extended, non-local, moves are allowed for.  In fact,
earlier Monte Carlo work \cite{Yoshino,Roters} had been hindered
by pathologies of the commonly used local move set which leads to
an infinite critical force for unbounded disorder.  Many workers
in the field \cite{Sneppen,TangLeschhornPRA,Buldyrev} had circumvented
these problems by rather considering cellular automaton models.
These models are very useful. However, it is normally impossible
to identify the differential equation which results in the continuum
limit.  In contrast, the Monte Carlo dynamics is derived from an
energy function, and it   satisfies  detailed balance.  We developed
a method which finds with great ease the critical string, i. e.
the string which is pinned at the critical force $f_c$.  In
\cite{RossoKrauth}, we  showed that the critical force and the
critical string are completely independent of all the details of
the (non-local) Monte Carlo algorithm.  Both from a conceptual and
a practical point of view, the situation is thus much better
controlled.  In this paper, we are concerned with the statistical
properties of critical strings for different functional forms of
the (short-range) elastic energy.

Specifically, we consider an elastic string  $h^t = \{
h_i^t\}_{i=0,\ldots,L}$ moving on a finite  lattice of size $L
\times M$ with periodic boundary conditions in both directions.
For concreteness, we take our random potential $V(i,j)$ at each
site to be made up of uncorrelated Gaussian variables with unit
variance.  The energy of a string $h^t$ in presence of an external
driving force $f$   is given by
\begin{equation}
E(h^t)=\sum_{i=1}^{L}\left\{ V(i,h_i^t)-f h_i^t
+E_{\mbox{el} }(\Delta_{i}^t)\right\}.
\label{energy}
\end{equation}
Here, the (short-range) elastic energy $E_{\mbox{el}}$ is a function
of  the local extension $\Delta_{i}^t= h^t_{i+1} - h^t_{i}$. Periodic
continuation is implied, such that, for large driving forces $f$,
the string $h^t$ keeps winding around the finite lattice.  The
Variant Monte Carlo method presented in \cite{RossoKrauth} allows
to compute the critical string $h^c$ for an arbitrary local convex
elasticity.

The string's roughness exponent $\zeta$,
in the thermodynamic limit, is defined by
\begin{equation}
(h^c_i - h^c_j)^2 \sim \mbox{const} | i-j | ^{2 \zeta}. 
\label{roughnessdef}  
\end{equation}
Following \cite{Jensen}, we obtain $\zeta$ 
by computing as a function of system size $L$ the mean square elongation
\begin{equation}
W^2(L): = \overline{
 < (h^c -   <h^c>)^2>
}.%end overbar 
\label{elongdef}
\end{equation}
In eq.(\ref{elongdef}), $<h^c> = \frac{1}{L}\sum_i h^c_i$, while
the overbar stands for an average over the disorder.

\begin{figure}
\centerline{ \psfig{figure=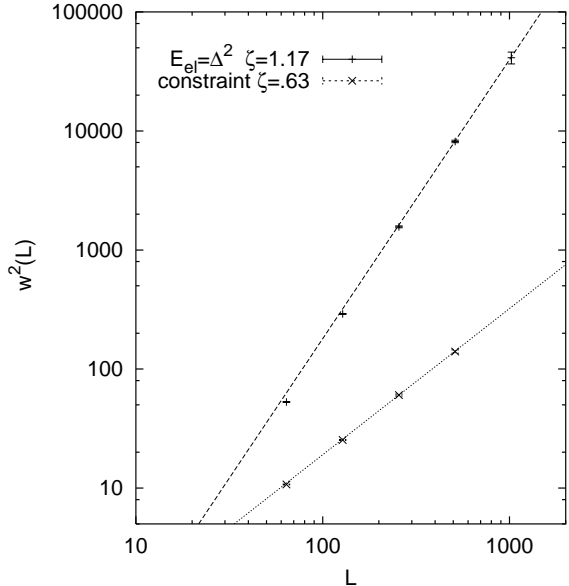,height=8.cm} }
\caption{ 
Mean square elongation $W^2(L)$ as a function of system size $L$
for a quadratic elastic energy $E_{\mbox{el} }(\Delta) = \Delta^2$
(upper curve) and for the model with  metric constraint $| \Delta|
\le 1$ and $a=0$ ({\em cf} eq.(\ref{metric})).  The interpolating lines
correspond to roughness exponents of $\zeta = 1.17$  and $0.63$,
respectively.  }
\label{figuresquare}
\end{figure}

As a first step, we show in figure \ref{figuresquare} the mean
square elongation $W^2(L)$ as a function of $L$ for a harmonic
elastic energy $E_{\mbox{el} }(\Delta) = \Delta^2$ for system sizes
$(L \times M)$ up to $(1000^2)$.  The data are very well fitted by
a straight line of slope $2.34$, which indicates that the roughness
exponent is $\zeta \simeq 1.17$.  It is now well understood that
a line with $\zeta > 1$  cannot represent a physical string
\cite{TangLeschhornDongcomm}. In fact, the thermodynamic
limit breaks down in this case, and the structure of the elastic string is
described by a size-dependent constant in eq.(\ref{roughnessdef})
$|h_i^c - h_j^c|^2 \sim \overline{\Delta^2(L)} | i - j| ^2$, with
a diverging mean square extension $\overline{\Delta^2(L)}$ in the
limit $L\rightarrow \infty$ \cite{footzetalarge}.

The same behavior has been observed in the  continuum limit $(i
\rightarrow x)$ of the lattice model eq.(\ref{energy}), where a
harmonic elastic energy yields a term $ \sim  \nabla^2
$ in the corresponding time evolution equation (quenched Edwards-Wilkinson
equation):
\begin{equation}
\frac{ \partial  }{\partial t } h(x,t) = f + \eta(x,h) + 2 a_1\nabla^2 h.
\label{Edwards_Wilkinson}     
\end{equation} 
Here, the  random force $\eta(x,h)$ is the negative derivative of
the random potential $V$ in eq.(\ref{energy}).  The
eq.(\ref{Edwards_Wilkinson}) has been much studied by analytical
methods \cite{Fisher,Nattermann,Chauve01} as well as by direct
numerical simulation \cite{Jensen,TangLeschhornDongcomm,Dong,Leschhorn}
and the existence of a roughness exponent in excess of one is now
well accepted.  From the numerical work, we expect a value of $\zeta
\simeq 1.15$ \cite{Jensen}, while Chauve {\em et al} \cite{Chauve01}
obtained $\zeta   \simeq 1.2 \pm 0.2$ from a two-loop functional
renormalization group calculation.
Our numerical  data in figure \ref{figuresquare} thus
establish perfect agreement of the lattice with the continuum
framework. 

The unphysical roughness of the elastic string described by the
quenched Edwards-Wilkinson equation has led many authors
\cite{Jensen,TangLeschhornDongcomm} to conclude that a $1-$dimensional
string necessarily develops overhangs and islands, which go beyond
a description by a single-valued function $h(x,t)$.  Others
\cite{Barabasi,KardarAniso} have attempted to introduce and to
justify nonlinear terms in the time evolution equation 
(\ref{Edwards_Wilkinson}) ({\em cf} below).

In this paper, we study elastic strings which {\em can} be described
by a function $h(x,t)$.  We pursue a approach based on the analysis
of the energy eq.(\ref{energy}) : We argue
that a diverging extension $\Delta$ is  contradictory with the
replacement of $E_{\mbox{el}}(\Delta)$ by the $\Delta^2$ term,
dominant only at small $| \Delta| $. Correspondingly, the quenched
Edwards-Wilkinson equation  stems from a small gradient expansion
\cite{Barabasi}, and is incompatible with this situation.  We find
that  $\zeta$ changes and becomes physical (and universal) if terms
beyond lowest order  in $\Delta$ are kept.  A single higher-order
contribution to the elastic energy is found to be important. It
generates a nonlinear piece in the corresponding continuum equation
of motion, whose origin has not been identified before.

We first consider an energy function, where the harmonic potential 
is cut off by a metric constraint:
\begin{equation}
E_{\mbox{el} } ( \Delta) = \left\{ \begin{array}{cc}
\infty & |\Delta| > 1 \\
a      & |\Delta| = 1 \\
0      & \Delta = 0 \\
\end{array}
\right. .
\label{metric} 
\end{equation}
By construction, the constraint in eq.(\ref{metric}) forces the
roughness exponent to be $ \zeta \le 1$.  Our data in the lower
part of figure \ref{figuresquare} evidence however a much more
interesting fact, namely an exponent $\zeta \simeq 0.63$.

This exponent coincides with the one obtained in many numerical
simulations of directed percolation depinning, as obtained with a
variety of cellular automata models.  Very interestingly, both the
cellular automaton (model $B$)  of Sneppen \cite{Sneppen} and the
rule proposed by Tang and Leschhorn \cite{TangLeschhornPRA} explicitly
provide for a local constraint which limits the value of $|\Delta|$,
as we can do in eq.(\ref{metric}). Notice that a whole  family of
Monte Carlo rules correspond to each energy function $E_{\mbox{el}
} $ and that the critical strings coincide for all its members.

\begin{figure}
\centerline{ \psfig{figure=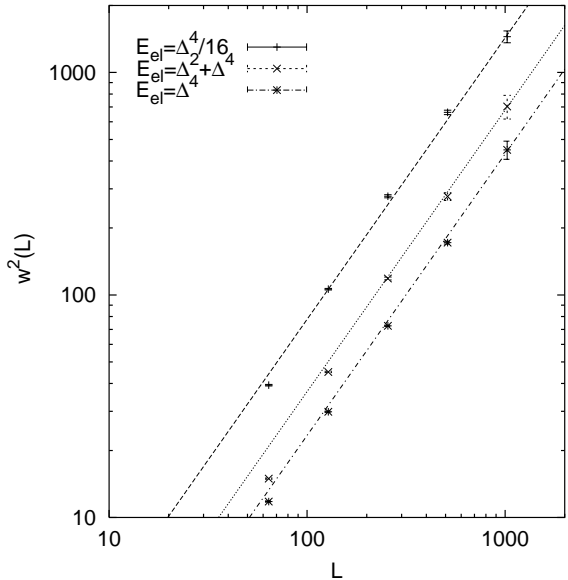,height=8.cm} }
\caption{
Mean square elongation $W^2(L)$ as a function of system size $L$
for elastic energies incorporating different quartic terms.
The interpolating lines correspond to roughness exponents
$\zeta = 0.63$, as for the metrically constrained curve in figure 
\ref{figuresquare}.}
\label{figurequartic}
\end{figure}

As mentioned in the beginning, we strive to keep full contact between
the Monte Carlo dynamics on the lattice  
and the continuum description. This can be achieved  more
easily if we consider smooth (but non-harmonic) functions 
$E_{\mbox{el} } ( \Delta)$ of which eq.(\ref{metric}) should come out
as a limiting case.
To do so, we note that the elastic energy must be a symmetric
function of the local extension $\Delta$.  We make the Ansatz to write  
it as a power series in
$\Delta^2$
\begin{equation}
E_{\mbox{el} }(\Delta) = a_1 \Delta^2 + a_2 \Delta^4 + \ldots,
\label{elasticenergy}
\end{equation}
where we assume all coefficients $a_i$ to be positive. In this way
we assure convexity of the function $E_{\mbox{el} }(\Delta)$, which
is a necessary prerequisite for our numerical algorithm.  By direct
computation of the roughness exponent in a variety of cases, we
have found the same exponent $\zeta \simeq 0.63$ whenever a term
of at least quartic order in $\Delta$   was present.  Consider, in
figure \ref{figurequartic}, our data for different choices  of the
coefficient $a_1$ and $a_2$ in eq.(\ref{elasticenergy}).  As can
be seen in the figure, the roughness exponent is in all cases $\zeta
\simeq 0.63$, a value which is not changed by  powers $\Delta^6$
and higher.  This value of $\zeta$ is physically acceptable, as it
yields a finite elastic energy per link of the critical string.
We conclude that the  elastic energy $E_{\mbox{el} } ( \Delta) $
contains only relevant terms of order $\Delta^2$ and  $\Delta^4$.
This insight allows us to directly consider the continuum limit.

In this  limit, the quenched Edwards-Wilkinson equation is modified 
by the negative gradient of the quartic term
\begin{equation}
\frac{ \partial  }{\partial t } h(x,t) = f + \eta(x,h) + 2 a_1\nabla^2 h 
+ 12 a_2 \nabla^2 h (\frac{ \partial h }{\partial x } )^2.
\label{Edwards_Wilkinson_quartic}     
\end{equation}

Our two observations:
\begin{itemize}
\item including a term $\Delta^4$ yields a
physically satisfactory roughness exponent $\zeta < 1$; 
\item powers
$\Delta^6$ and higher do not influence the value of $\zeta$; 
\end{itemize}
can now be interpreted in the light of the standard  hydrodynamic
scaling argument ({\em cf} \cite{Barabasi}).  This argument supposes
the string to be self-affine, so that a scale transformation $ x
\rightarrow b x$ corresponds to   a transformation $h \rightarrow
b^{\zeta} h$. In the hydrodynamic limit ($b \rightarrow \infty$),
higher derivatives (corresponding to higher-order terms in the
elastic energy eq.(\ref{elasticenergy})) are negligible compared
to lower ones if the roughness exponent satisfies $\zeta < 1$.
Notice that the hydrodynamic scaling argument is self-consistent:
The roughness exponent has to satisfy $\zeta < 1$ even  {\em after}
dropping higher order terms. The naive application of the argument
would lead us to neglect also the last term in
eq.(\ref{Edwards_Wilkinson_quartic}). This would bring us back to
the quenched Edwards-Wilkinson equation, which describes a string
which is not self-affine.

The discrepancy between the values of the roughness exponents
$\zeta\simeq 1.17$ and $\zeta \simeq 0.63$ has usually been
interpreted as due to the presence of two distinct universality
classes for interface growth in disordered environments \cite{Barabasi},
one belonging to the quenched Edwards-Wilkinson equation
(eq.(\ref{Edwards_Wilkinson}))
\cite{TangLeschhornDongcomm,Fisher,Nattermann,Chauve01,Dong}, the
other represented by the cellular automata models
\cite{Sneppen,TangLeschhornPRA,Buldyrev} as well as by experiments
on directed percolation depinning \cite{Buldyrev}.  We would argue
that $\zeta=1.17$ is a specific result, valid for the harmonic
potential only, and strongly dependent on the behavior
$E_{\mbox{el}}(\Delta)$ for $| \Delta|  \rightarrow \infty$.

Numerical work by Amaral {\em et al} \cite{Amaral} had detected
the presence of a nonlinear relevant term  in cellular automata
that was initially believed to be caused by the $\lambda  (\frac{
\partial h }{\partial x } )^2 $ term in the quenched  KPZ  equation
\cite{KPZ}
\begin{equation}
\frac{ \partial  }{\partial t } h(x,t) = f + \eta(x,h) + \nu \nabla^2 h
+ \lambda/2  (\frac{ \partial h }{\partial x } )^2.
\label{KPZ}
\end{equation}     
However, in the KPZ equation this term is of kinematic origin and
vanishes at the depinning threshold \cite{Fisher,Kardar,Parisi},
where the critical string is a purely static object.  It was later
believed \cite{KardarAniso,Kardar} that anisotropies in the disorder
distribution might be responsible for such a nonlinear term.
 
In our system, an anisotropy is clearly absent. Rather, we have
shown here that a roughness exponent $\zeta \simeq 0.63$ is naturally
generated either by a higher order term in the elastic energy ({\em cf}
eq.(\ref{elasticenergy})) or, equivalently, by a metric constraint
as in eq.(\ref{metric}). Such a constraint is also present in
cellular automata models. Without it, we generate an unphysical
elastic string with $\zeta \simeq 1.17$.  The nonlinear term we
introduced into eq.(\ref{Edwards_Wilkinson_quartic}) is the most
relevant one which can be generated at the microscopic level within
the static description of an energy function eq.(\ref{energy}).
Note that such a description is sufficient below the
depinning threshold, where viscous or inertial effects are naturally
absent.  We suspect that the nonlinear term may generate other,
more relevant, terms upon coarse graining (for related work, {\em cf}
\cite{Braunstein}).  This point will certainly have to be studied
analytically, and goes beyond the scope of this paper.

The quenched Edwards-Wilkinson equation has also been studied for
the driven motion of $d$-dimensional interfaces in a $d+1$ dimensional
target space.  The $d=4$ dimensional exact, stable, solution has
been used as a starting point for perturbative renormalization
group calculations in dimension $d=4-\epsilon$
\cite{Fisher,Nattermann,Chauve01}. The fact that, in $d=1$, the
roughness exponent $\zeta >  1$ seems to indicate that, by continuity,
our  nonlinear term will be present even in higher dimensions.
There are indications of at least logarithmic divergencies even in
$d=2$ \cite{Jensen}. Clearly, much effort will still be needed in
order to completely understand this system.

In conclusion, we have studied in this paper the effect of the
elastic energy on the statistical properties of a string at the
depinning threshold. For a harmonic energy $E_{\mbox{el}} \sim
\Delta^2$, we find a roughness exponent $\zeta$ familiar from
numerical and analytical work on the quenched Edwards-Wilkinson
equation. This exponent indicates that the string is unphysical,
as the mean square extension of the string $\Delta^2$ diverges in
the thermodynamic limit.  We then studied different elastic energies,
either by cutting off the harmonic function with a metric constraint
on $| \Delta| $   or by using a more general (differentiable) Ansatz
for $E_{\mbox{el}} (\Delta)$. In both cases we regularized the
string and obtained a physically sound roughness exponent $\zeta
\simeq 0.63$. The lowest order addition to the elastic energy, of
quartic order in $\Delta$, generates a nonlinear piece in the
corresponding continuum time evolution equation.

Although we do not know at present how this new nonlinear term
(which preserves te symmetry $h \rightarrow -h$) will behave under
coarse graining, we can be quite sure that it is correct on a
microscopic level and gives a consistent thermodynamic description
of a driven elastic string at the depinning threshold.  It will
generate a roughness exponent of $\zeta \simeq 0.63$ in the continuum.

Finally, we note that the Monte Carlo methods introduced in
\cite{RossoKrauth} have allowed to obtain the critical string even
for large systems, and for general convex energy functions. This
yields the  completely transparent transition between the two
situations with $\zeta = 1.17$ and $\zeta \simeq 0.63$, of which
only the latter is universal and has physical relevance \cite{footprog}.

Acknowledgments: During the course of this work we have benefited
from very valuable input by  P. Le Doussal. We also thank A. A.
Middleton, L. Santen, and J. Vannimenus for helpful discussions.

%\bibliographystyle{prsty}
%\bibliography{refbec,refman}

\begin{thebibliography}{10}
\bibitem{footzetalarge} The value of $\zeta$ can still be measured
using eq.(\ref{elongdef}) \cite{Jensen}.
\bibitem{footprog}  The Fortran program used to compute the critical elastic
string is available from the authors.
\end{thebibliography}

\end{multicols}

\end{document}